\documentclass[doublecol]{epl2}

\title{Trapped Bose gases with large positive scattering length}

\author{M.~Th{\o}gersen, D.V.~Fedorov, and A.S.~Jensen}

\institute{IFA, Aarhus University, 8000 Aarhus C, Denmark}

\pacs{}{03.75.Hh, 03.75.Nt, 21.45.+v}

\abstract{ We calculate the energy and the condensate fraction of a
system of trapped bosons interacting via a short-range two-body potential
with positive scattering length.  The potential is attractive and has a
two-body bound state.  When the scattering length is small compared to
the trap length the system is model independent: all potential models --
attractive, repulsive and zero-range -- provide similar results.  When the
scattering length is large the attractive model differs qualitatively
from the repulsive and zero-range models. In this regime the system with
attractive potential becomes independent of the scattering length, with
both the energy and the condensate fraction converging towards finite
constants. }

\begin{document}

\maketitle

\section{Introduction}

The density of trapped cold gases is generally low under typical
experimental conditions~\cite{court}, such that the parameter $nb^{3}$
is small, where $n$ is the particle density and $b$ is the range of the
inter-particle potential. In other words the typical distance between
particles is much larger than the range of the potential, and the typical
relative momentum between particles is much smaller than the inverse
range of the potential.

In this regime the system of particles exhibits universality (also called
model independence or shape independence) \cite{braaten}. The system is
not sensitive to the details of the potential and the properties of the
system are essentially determined by only few low-energy parameters of the
potential. Very different interaction models then provide quantitatively
similar results as soon as the low-energy parameters are the same.

In a two-body system the universality is manifested in the well known
effective range expansion, where the low-energy $s$-wave phase shift
$\delta$ is determined by only two parameters, the scattering length $a$
and the effective range $r_e$,
\begin{equation}\label{eq:effr}
k\cot\delta = -\frac{1}{a}+\frac{1}{2}r_ek^2+O\left(r_e^3k^4\right),
\end{equation}
where $k$ is the relative momentum between particles. The effective range
$r_e$ is typically of the order of the range of the potential while the
scattering length can vary greatly.

In three-body systems the universality manifests itself in the
Thomas~\cite{thomas} and Efimov~\cite{efimov} effects, the Phillips
line~\cite{phillips}, and other low-energy phenomena~\cite{nielsen},
also in two dimensions~\cite{2d}.

Cold gases also exhibit universality: in the dilute limit their
properties, in particular the energy per particle, are independent on
the shape of the inter-particle potential and are determined by the
scattering length alone. This universality is customarily employed by
using the zero-range (pseudo-)potential for theoretical descriptions
of the Bose-Einstein condensates. The zero-range potential has only
one parameter, the scattering length. Combined with the Hartree-Fock
product wave-function the zero-range potential model is known as the
Gross-Pitaevskii equation~\cite{BEC}.

The scattering length in a trapped atomic gas can be tuned to
an essentially arbitrary value using the technique of Feshbach
resonances~\cite{court}. This gives a possibility to investigate trapped
systems with very large scattering lengths. In this case the first term
in the effective range expansion eq.~(\ref{eq:effr}) vanishes and the
system becomes sensitive to the effective range of the potential.

The limits of the zero-range model have been tested by numerical
calculations with finite-range potentials.  In particular, for the large
positive scattering length repulsive potentials have been employed within
Monte-Carlo methods~\cite{blume01,gbc,dubois01}.  These investigations
showed that as the scattering length is increased the energy of the Bose
gas with a repulsive potential exceeds the zero-range predictions
and the condensate fraction becomes considerably depleted.

However, repulsive potentials have a problem when modelling an
increasingly large positive scattering length: the effective range of the
potential has to be increased essentially linearly with the scattering
length. This does not seem to match the experimental conditions where
the scattering length is adjusted by tuning the atomic resonances in an
external magnetic field. The range of the inter-atomic interaction is
then left essentially unchanged.

Instead, an attractive finite-range potential might be a more realistic
interaction model for descriptions of trapped Bose gases with Feshbach
resonances.  Indeed, with an attractive potential an arbitrary large
positive scattering length can be achieved by fine-tuning the energy
of the bound two-body state, while maintaining the given realistic
effective range.

However, attractive potentials with bound states bring in a major
complication for numerical calculations: a large number of many-body
self-bound negative-energy states appears in the system and the
condensate state in the trap becomes a highly excited state.

For a homogeneous Bose gas an approximate Jastrow-type wave-function was
employed where the pair-correlation function was essentially a solution
of the two-body equation~\cite{cowell,gbc}.  In contrast we propose a
direct numerical diagonalisation of the many-body Hamiltonian where the
condensate state of trapped bosons appears as a many-body excited state
which is automatically orthogonal to all the self-bound negative-energy
states.

The purpose of this paper is to investigate the energy and the condensate
fraction of a system of trapped bosons with attractive potentials as
function of the scattering length and the number of bosons, and compare
the results with the zero-range and repulsive model.

\section{System and numerical techniques}

\subsection{The system}

We consider a system of $N$ identical bosons with mass $m$ and coordinates
$\mathbf{r}_{i}$,  $i=1,\dots,N$, in a spherical harmonic trap with
frequency $\omega$. The Hamiltonian of the system is given by
\begin{equation}\label{eq:ham}
H=-\frac{\hbar^{2}}{2m}
\sum_{i=1}^{N}\frac{\partial^2}{\partial\mathbf{r}_{i}^{2}}
+\sum_{i<j}V(\left|\mathbf{r}_{i}-\mathbf{r}_{j}\right|)
+\frac{m\omega^{2}}{2}\sum_{i=1}^{N}r_{i}^{2}\;,\label{eq:h}
\end{equation}
where the system parameters are taken from \cite{blume01}:
$m=86.909$amu, $\omega=2\pi\times77.87$Hz, the trap length
$b_t=\sqrt{\hbar/(m\omega)}=23095$au.

\subsection{Two-body potentials}

The zero-range potential model,
\begin{equation}\label{eq:zrp}
V_{ZR}(r)=\frac{4\pi\hbar^{2}a}{m}\delta(r)\;,
\end{equation}
has only one length parameter, $a$. For dilute bosonic systems this
parameter is customarily chosen to be equal the inter-atomic scattering
length.  The zero-range potential provides then the correct low-energy
scattering amplitude in the first order Born approximation. The zero-range
potential can only be used with an appropriate non-correlated functional
space~\cite{uncorr}, like the Hartree-Fock product wave-functions. In the
latter case it leads to the famous Gross-Pitaevskii equation~\cite{BEC}.

For the finite-range potential model we use a Gaussian,
\begin{equation}\label{eq:gau}
V(r)=V_{0}\exp(-{r^{2}\over b^{2}})\;,
\end{equation}
with the range $b=11.65$au and a varied negative strength $V_0$. The
variation of the strength is limited to the region where the potential
provides exactly one two-body bound state and a positive scattering
length.

\subsection{Stochastic variational method}

The wave-function of the system is represented as a linear combination
of $K$ basis-functions taken in the form of symmetrised correlated Gaussian,
\begin{equation}
\Psi=
\hat{S}\sum_{k=1}^{K}C_{k} \exp
\left(
-\frac{1}{2}\sum_{i<j}^{N}\alpha_{ij}^{(k)}
(\mathbf{r}_{i}-\mathbf{r}_{j})^2
\right)
\;,\label{eq:psi-full}
\end{equation}
where $\hat{S}$
is the symmetrisation operator, and $C_{k}$ and $\alpha_{ij}^{(k)}$ are
variational parameters. The linear parameters $C_{k}$ are determined by
an ordinary diagonalisation of the Hamiltonian eq.~(\ref{eq:h}) while the
non-linear parameters $\alpha_{ij}^{(k)}$ are optimised stochastically
by random sampling~\cite{varga,hansh} from a region that covers the
distances from $b$ to $b_t$. The center-of-mass motion is assumed to be
in the oscillator's ground state.

The zero-range potential eq.~(\ref{eq:zrp}) requires an uncorrelated
wave-function which we chose in the form of the linear
combination of the hyper-radial basis-functions,
\begin{equation}\label{eq:psi-hr}
\Psi_{\rho}=
\sum_{k=1}^{K}C_{k}
\exp\left(-\frac{1}{2}\alpha^{(k)}\rho^{2}\right)
\;,\label{eq:psi-rho}
\end{equation}
where $\rho^{2}=\sum_{i=1}^{N}r_{i}^{2}$ is the hyper-radius of the
system. This function is totally symmetric and thus does not require the
symmetrisation operator ${\hat S}$. The zero-range potential with the
hyper-radial variational wave-function eq.~(\ref{eq:psi-rho}) provides
results similar to Gross-Pitaevskii equation \cite{hansh}.

The calculation of a highly excited state with the fully correlated
basis eq.~(\ref{eq:psi-full}) is a difficult numerical task and is only
possible for relatively small number of particles.

However for a typical system of trapped atoms even when the scattering
length is large the density of the system remains small, $nb^{3}\ll1$,
and one can assume that only binary collisions play a significant
role in the system's dynamics.  In this approximation the variational
wave-function can be simplified by only allowing two-body
correlations in the basis-functions,
\begin{eqnarray}
\Psi_{2b} &=& \hat{S}\sum_{k=1}^{K}C_{k} \nonumber \\
&\times&\exp\left( -\frac{1}{2}\alpha^{(k)}\rho^{2}
-\frac{1}{2}\beta^{(k)}({\mathbf
r}_{1}-{\mathbf r}_{2})^{2}\right) \;,
\label{eq:psi-2b}
\end{eqnarray}
where
$\alpha^{(k)}$ and $\beta^{(k)}$ are the nonlinear parameters. The
symmetrisation of this function can be done analytically \cite{hansh}
which greatly simplifies the numerical calculations.

During the calculation of a given system the number of Gaussian in the
basis is increased and the stochastic optimisation is carried out until
the number of negative energy states and the energy of the lowest state
with positive energy is converged.  The convergence within four digits
typically requires about $5\times 10^2$ Gaussian and about $10^5$
random trials per nonlinear parameter. The stochastic optimisation
algorithm is easily parallelisable with close to linear scalability.

\subsection{Condensate fraction}
We first calculate the one-body density matrix defined as
\begin{eqnarray}
n(\mathbf{r},\mathbf{r}^{\prime})&=&
\int d\mathbf{r}_{2}\dots d\mathbf{r}_{N} \nonumber \\
&\times&\Psi^{\star}(\mathbf{r},\mathbf{r}_{2},\dots,\mathbf{r}_{N})
\Psi(\mathbf{r}^{\prime},\mathbf{r}_{2},\dots,\mathbf{r}_{N})
\;,\label{eq:nrr}\end{eqnarray}
where $\mathbf{r}_i$ are the coordinates of the atoms measured from
the center of the trap.  The density matrix is normalised to unity
independent of the number of particles.

The density matrix is then diagonalised, meaning that its single-particle
eigenfunctions $\chi_{i}(\mathbf{r})$ and the corresponding eigenvalues
$\lambda_{i}$ are calculated,
\begin{equation}
\int d\mathbf{r}^{\prime}n(\mathbf{r},\mathbf{r}^{\prime})
\chi_{i}(\mathbf{r}^{\prime})
=\lambda_{i}\chi_{i}(\mathbf{r})
\;.\label{eq:nchi}\end{equation}
The condensate fraction, $\lambda_{0}$, is then defined as the largest
eigenvalue.

\subsection{Bose-Einstein condensate state}

\begin{figure}[htb]
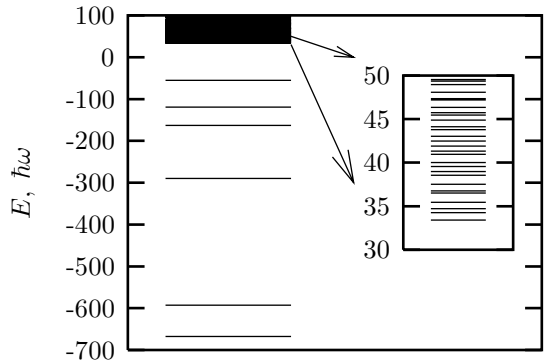

\centerline{\input fig-spe}
\caption{The typical spectrum (in the vicinity of zero energy) of a system
of $N=20$ bosons in an oscillator trap eq.~(\ref{eq:ham}) interacting
via an attractive Gaussian two-body potential eq.~(\ref{eq:gau}) with
one bound state and a positive scattering length. The inset shows
the beginning of the quasi-continuum  spectrum.  } \label{fig:spe}
\end{figure}

For repulsive potential models it is simply the ground state of the trapped
many-boson system that is identified as the Bose-Einstein condensate state
(BEC-state).

With deep attractive two-body potentials, however, the many-body system
in a trap has a large number of self-bound negative-energy states and
identification of the BEC-state is not obvious.

The typical spectrum of a trapped many-body system with attractive
potentials is shown on fig.~\ref{fig:spe}. The system has a number
of deeply bound states with negative energies and then a positive
quasi-continuum spectrum which starts at about $\frac{3N}{2}\hbar\omega$
and has the characteristic distance between levels of the order
$\hbar\omega\ll \hbar^2/(2mb)^2$.

Apparently the BEC-state should then be the lowest state of the
quasi-continuum spectrum or, equivalently, the lowest state with positive
energy.

\begin{figure}[htb]
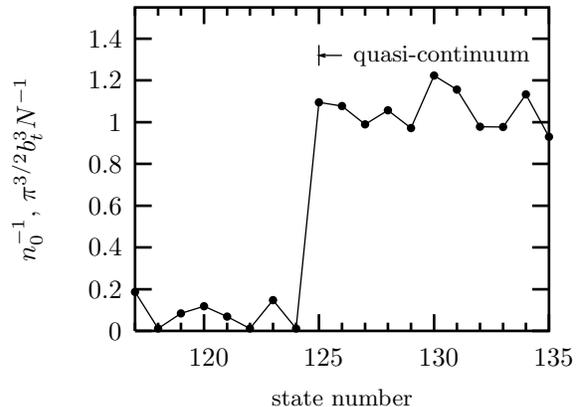

\centerline{ \input fig-v.tex }
\caption{The inverse central density $n_0^{-1}$ in oscillator units for
a system of trapped bosons from fig.~\ref{fig:spe} as function of the
state number. Shown are only the states in the vicinity of the beginning
of the quasi-continuum spectrum.}
\label{fig:v}
\end{figure}

\begin{figure}[htb]
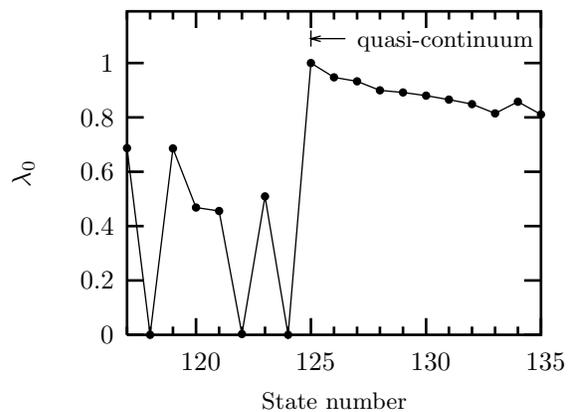

\centerline{ \input fig-cfs.tex }
\caption{The condensate fraction of a system of trapped bosons from
fig.~\ref{fig:spe} as function of the state number. Shown are only the
states in the vicinity of the beginning of the quasi-continuum spectrum.}
\label{fig:cfs}
\end{figure}

To verify this assumption we calculate the central density, $n_0$,
of the system for the negative- and positive-energy states around zero
energy. The results are shown on fig.~\ref{fig:v} in the form of the
inverse central density (the volume per particle) $n_0^{-1}$.  In the
BEC-state the atoms should occupy the whole volume of the trap and thus
the volume per particle should be close to one (in the correspondingly
scaled oscillator units). The states with negative energy are self-bound
states with much higher density and thus much smaller volume per
particle. And indeed that is what the figure shows -- a sharp increase
in the volume per particle from a small value to about unity exactly at
the lowest state with positive energy where the quasi-continuum starts.

Another test is the very condensate fraction, shown for several states
around zero energy on fig.~\ref{fig:cfs}. The self-bound states with
negative energy must have smaller condensate fraction compared to the
BEC-state, and the excitations from the BEC-state must gradually deplete
the condensate fraction. Apparently this is what is seen on the figure
-- a sharp increase of the condensate fraction to about 100\% at the lowest
state with positive energy with the subsequent gradual depletion.

We have thus verified that in the case of attractive potentials the
BEC-state of a system of trapped bosons is the lowest state with positive
energy.

\subsection{Accuracy of the two-body correlated basis}

\begin{table}[htb]
\caption{
The energies in units of $\hbar\omega$ for the BEC-state of a system of
$N$ bosons in a harmonic trap eq.~(\ref{eq:ham}) for different interaction
models with the same scattering length of 100~au. For the Gross-Pitaevskii
(GP), hard-spheres (HS), and zero-range (ZR) models the BEC-state is
the ground state; for the attractive model (A) the BEC-state is the lowest
state with positive energy. The Gross-Pitaevskii and hard-spheres data
are taken from \cite{blume01}. The attractive model employed the
two-body correlated basis eq.~(\ref{eq:psi-2b}). }

\label{tab:blume}
\begin{center}
\begin{tabular}{|c|c|c|c|c|}
\hline 
$N$ & GP & HS & ZR & A
\tabularnewline
\hline
\hline 
3   & 4.51032    & 4.51036(2)  & 4.5103        & 4.510
\tabularnewline\hline 
5   & 7.53432    & 7.53443(4)  & 7.5342        & 7.534
\tabularnewline\hline 
10  & 15.1534    & 15.1537(2)  & 15.1533       & 15.154
\tabularnewline \hline 
20  & 30.638     & 30.640(1)   & 30.6394       & 30.640
\tabularnewline \hline
\end{tabular}
\end{center}
\end{table}

In ref.~\cite{blume01} the energies of several low-density systems of
trapped bosons were calculated using the Gross-Pitaevskii and repulsive
hard-sphere models with the same "natural" scattering length of 100~au.
In this regime the systems exhibit universality and the energies
calculated in both models were very close.

To test the accuracy of our two-body correlated basis
eq.~(\ref{eq:psi-2b}) (which is expected to be a good approximation in
the low-density regime) as well as the identification of the BEC-state
for attractive potentials we consider the same systems with the same
scattering length but with the attractive potential eq.~(\ref{eq:gau}) and
calculate the energy of the BEC-state according to our prescription.  The
BEC-state is now an excited state and is identified in the calculations
as the lowest state with positive energy.  We also calculate the energies
for the zero-range potential model eq.~(\ref{eq:zrp}) with hyper-radial
trial wave-function~eq.~(\ref{eq:psi-hr}).

The results are given in table~\ref{tab:blume}. As expected, these
low-density systems with relatively short scattering length exhibit
universality as all potential models give essentially the same results. We
conclude that we do a correct identification of the BEC-state and that
that the two-body correlated basis has an adequate accuracy.

\begin{table}[htb]
\caption{\label{table:ft} The energies, in units of $\hbar\omega$, of the
lowest state with positive energy for a system of 4 bosons in a harmonic
trap eq.~(\ref{eq:ham}). The bosons interact via an attractive Gaussian
potential eq.~(\ref{eq:gau}) with the strength $V_0$ and the scattering
length $a$. The results from fully-correlated basis eq.~(\ref{eq:psi-full})
and from the two-body correlated basis eq.~(\ref{eq:psi-2b}) are designated
correspondingly $E$(full) and $E$(2b).}
\begin{center}
\begin{tabular}{|c|c|c|c|}
\hline
$V_0$, au & $a$, au & $E$(2b) & $E$(full)
\tabularnewline\hline\hline
-1.400e-7 &  119.4  &   6.025    &    6.025
\tabularnewline\hline
-1.300e-7 &  327.0  &   6.067    &    6.067
\tabularnewline\hline
-1.290e-7 &  402.4  &   6.083    &    6.083
\tabularnewline\hline
-1.280e-7 &  525.4  &   6.108    &    6.108
\tabularnewline\hline
-1.270e-7 &  761.0  &   6.155    &    6.156
\tabularnewline\hline
-1.260e-7 &  1400   &   6.282    &    6.283
\tabularnewline\hline
-1.255e-7 &  2430   &   6.478    &    6.481
\tabularnewline\hline
-1.252e-7 &  4370   &   6.818    &    6.848
\tabularnewline\hline
-1.251e-7 &  5962   &   7.059    &    7.112
\tabularnewline\hline
\end{tabular}
\end{center}
\end{table}

To check the accuracy of the two-body correlated basis also for
large scattering lengths we perform a test calculation for 4 particles
with fully correlated and with two-body correlated basis for vastly
different scattering lengths. The results are given in table~\ref{table:ft}.

Although the accuracy of the two-body correlated basis decreases somewhat
with the increase of the scattering length, yet the relative accuracy
is better than 1\% even for exceedingly large scattering lengths.

\section{Results}
\subsection{Energy}

\begin{figure}
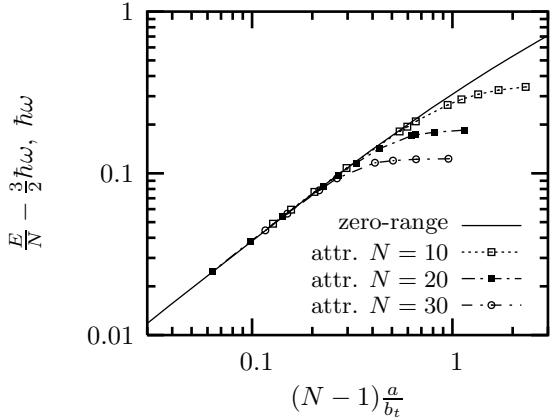

\centerline{ \input fig-e.tex }
\caption{The energy per particle $E/N$ as function of the scattering
length $a$ for a BEC-state of a system of $N$ identical bosons
in a harmonic trap eq.~(\ref{eq:ham}) for the zero-range model
eq.~(\ref{eq:zrp}) (solid line), and attractive potential model
eq.~(\ref{eq:gau}) (broken lines).  The zero-range results are
$N$-independent in this parametrization.  }
\label{fig:e} \end{figure}

Fig.~\ref{fig:e} shows the energy per particle for different $N$
and $a$ for two interaction models, zero-range potential and attractive
potential. For small scattering lengths the different potential models
give the same universal results -- the system is model independent. For
larger scattering lengths the energies from the attractive model are
systematically below the zero-range model, quite unlike the repulsive
model which goes above the zero-range model~\cite{blume01,dubois01}. For very
large scattering lengths the attractive model, unlike the zero-range
and repulsive models, becomes insensitive to the scattering length and
the energies converge to a constant. This is consistent with the
Jastrow-type approximation of ref.~\cite{cowell}.

In the regime close to the two-body threshold, where the scattering length
is large, an arbitrary large change in the scattering length needs only
an infinitesimally small change of the depth of the attractive potential
(see table~\ref{table:ft}). Therefore when the scattering length is larger
than the trap length it ceases to be a physical length scale for the
system which is then only subjected to a very small change of the depth
of the two-body potential which rapidly converges towards the two-body
threshold value. Since the external oscillator potential turns all
continuum states into discrete states all singularities due to various
thresholds are removed. Then clearly an infinitesimally small change
in the potential, despite the large change in the scattering length,
only leads to a linearly infinitesimal change in the system which then
becomes independent on the scattering length.

On the contrary for the finite-range repulsive potential models an
increase of the scattering length needs an almost proportional increase
in the effective range of the potential. Thus the system never ceases
to depend on the scattering length.

\begin{figure}[htb]
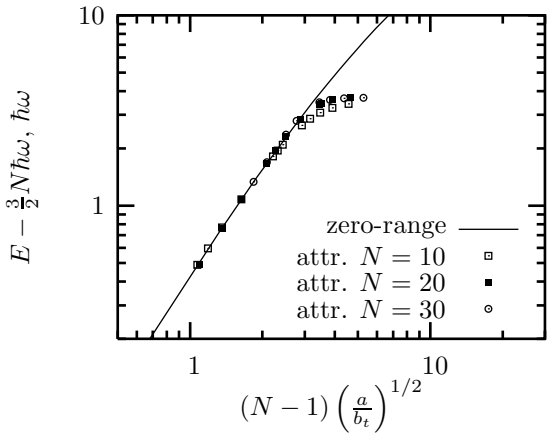

\centerline{ \input fig-e2.tex }
\caption{The same as is in fig.~\ref{fig:e} but plotted in a different
parametrisation.
\label{fig:e2} }
\end{figure}

If we plot the data using a different parametrisation, namely $E$ as
function of $(N-1)\left({a\over b_{t}}\right)^{1/2}$, the energy data
points seem to follow a "universal" curve as shown on fig.~\ref{fig:e2}.

\subsection{Condensate fraction}

\begin{figure}[hbt]
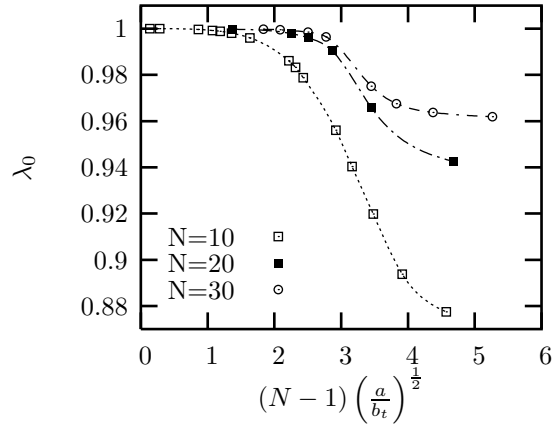

\centerline{\input fig-cf.tex}
\caption{The condensate fraction of a BEC-state of a system of $N$
identical bosons
in a harmonic trap eq.~(\ref{eq:ham}) with attractive inter-particle potential
eq.~(\ref{eq:gau}) as function of the scattering length $a$.}
\label{fig:cf}
\end{figure}

Our results for the condensate fraction of a system of trapped identical
bosons in a BEC-state is shown on fig.~\ref{fig:cf}. For small scattering
lengths the system is 100\% condensate. Then when $(N-1)\left({a\over
b_{t}}\right)^{1/2}$ is about 3 the condensate fraction rapidly drops a
few percent before stabilising again. Although we could not reach further
due to numerical convergence problems, we expect that the condensate
fraction will not appreciably change with the further increase of the
scattering length.
This seems consistent with the energies being stagnated when approaching
the two-body threshold.

This behavior is qualitatively different from what happens
within the repulsive models. In the hard-spheres Monte-Carlo
simulations~\cite{dubois01} the condensate fraction starts to deviate
from 100\% much later, when ${a\over b_{t}}$ is of the order of 0.1,
and with further increase of the scattering length the condensate is
completely quenched.

Again our results are different from the Jastrow-type
approximation~\cite{cowell}, where the condensate also becomes fully
quenched. However the authors of ref.~\cite{cowell} believe that their
estimates for the condensate fraction are "rather crude".

We have to note also that our wave-function includes only two-body
correlations which may lead to an over-estimate of the condensate
fraction.

\section{Conclusions}

We have calculated the energy and the condensate fraction of a system of
$N$ bosons in a harmonic trap as function of the number of bosons and
the scattering length $a$. Specifically we considered the regime where
the scattering length is positive and not small compared to the trap
length. The positive scattering length is modeled using an attractive
two-body potential with a bound two-body state. The many-body system
then has a large number of negative-energy self-bound states and the
condensate in the trap is identified as the lowest excited state with
positive energy.

When the scattering length is small compared to the trap length the
system shows model independence (universality) -- the results from the
attractive potential model are very close to those from the zero-range
and the repulsive potential models.

In the limit of large scattering length the system becomes independent
of the scattering length, contrary to the zero-range and the repulsive
models.

The condensate fraction decreases with the scattering length and reaches
a finite constant at large scattering lengths contrary to the repulsive
models where it reaches zero in this limit.

For the attractive potentials the energy per particle of the system of
trapped bosons follows a universal curve.

\acknowledgements
Numerous fruitful discussions with H.H.~Soerensen and T.~Kjaergaard are
acknowledged.

\end{document}